\documentclass{llncs}
\usepackage{booktabs}
\usepackage{caption}
\usepackage{subcaption}
\usepackage{multirow}
\usepackage{graphicx}
\usepackage{amsmath}
\usepackage{algorithm}
\usepackage[noend]{algpseudocode}

\begin{document}

\title{Improving Ads-Profitability Using Traffic-Fingerprints}
\author{Adam Gabriel Dobrakowski\inst{1} \and 
Andrzej Pacuk\inst{1} \and
Piotr Sankowski\inst{1,2,3} \and
Marcin Mucha\inst{3} \and
Paweł Brach\inst{4}}
\institute{MIM Solutions, Poland
\email{\{adam.dobrakowski,andrzej.pacuk,piotr.sankowski\}@mim-solutions.ai}
\and IDEAS NCBR, Poland
\and University of Warsaw, Poland
\email{mucha@mimuw.edu.pl}
\and HitDuck, Poland
\email{pbrach@hitduck.com}
}

\maketitle
\begin{abstract}
This paper introduces the concept of {\em traffic-fingerprints}, i.e., normalized 24-dimensional vectors representing a distribution of daily traffic on a web page. 
Using k-means clustering we show that similarity of traffic-fingerprints 
is related to the similarity of profitability time patterns for ads shown on these pages. In other words, these fingerprints are correlated with the conversions rates, thus allowing us to 
argue about conversion rates on pages with negligible traffic. 
By blocking or unblocking whole clusters of pages we were able to 
increase the revenue of online campaigns by more than 50\%.
\end{abstract}

\section{Introduction}

Internet becomes more and more anonymous which poses new challenges in front of advertisers. On the positive side, we are currently on the way to guarantee users the comfort of not being tracked all the time, e.g., via cookies. On the negative side, it is becoming harder to make sure that a user is interested in the content of an ad. However, from the point of view of the advertisers, not only users are being represented by unidentifiable numbers, but also web pages or even ad placements. In particular, this is the case for many advertising platforms that connect advertisers to web pages that want to host advertisements.
The key function of advertising networks is an aggregation of ad supply from many different web publishers and matching it with advertiser’s demand. Many ad networks use publishers’ domains masking to hide the actual domain name, publisher or placement identifier. For instance, MGID ({\tt www.mgid.com}) is an advertising platform that identifies a page (or an ad slot) by a UID ({\em unique traffic source ID}) that provides no information about its content, users visiting it nor impressions. Hence, it is hard to tell whether our ad will be well placed on a given page. This problem is further amplified by the fact that MGID connects many small publishers which attract a limited number of clicks from our campaigns, e.g., few clicks per day. In other words, we cannot directly estimate the performance of our campaign for a given UID in any statistically reasonable way as the conversion rates (CRs) are of the order of tenths of a percent. Moreover, each day new pages appear. Specifically, when we try to block some underperforming pages the networks will deliver traffic via new sources. 

Agencies and media operators are responsible for their clients’ marketing campaigns. They have to rely on measurement to optimize many campaigns’ metrics such as ROI, CR or profit. However, manually going through multiple campaign statistics can be very ineffective. Moreover, it is not possible to ensure by the human operators that the campaigns remain profitable on all domains, as this requires combining knowledge of many different statistics.

In this paper, we propose a novel method for enriching existing performance marketing software with a domain blocking algorithm to improve campaigns’ related KPIs. We define {\em traffic-fingerprints} for UIDs which are defined as normalized 24-dimensional vectors representing the number of clicks collected in each hour of the day over the history of interaction with each page. 
The traffic-fingerprints are divided into groups using k-means clustering. Then we treat pages in each cluster jointly, i.e., we turn all of them on or off during specific hours. This approach allowed us to increase the profit of our MGID-campaigns by 53\%. These proof-of-concept results were obtained via A/B tests on a system that was enriched with the above blocking algorithm.

\paragraph{Related Work}

Understanding and modeling web traffic is a topic that has attracted research for years already~\cite{10.1145/2068816.2068845,Thompson1997WideareaIT}. Moreover, unsupervised learning techniques like clustering have been widely used for analysing web traffic. However, such analyses have been typically done to create patterns for user behaviour. For example, in~\cite{NASIR2021101687} users are being segmented based on their social media advertising perceptions which reveals groups of users more susceptible to such ads. Another context for clustering users based on behaviour is web page recommendation~\cite{9200312} or product recommendation~\cite{Vanessa2018ContextualMB}. Although these papers use contextual data, some approaches create user profiles solely based on click-stream data~\cite{Singh2021AnEC,click-stream,SU20151,inbook,recommendation,10.1093/llc/fqx030}. These last approaches differ from ours in a two-fold way. First, we cluster pages and not users. Second, we use traffic-fingerprints, whereas these papers study the whole user click-streams, which in our context are not available. We note, however, that some of these papers studied behaviour patterns with respect to time~\cite{10.1093/llc/fqx030,Vanessa2018ContextualMB} and have revealed that some user 
profiles are active over a specified period of days, e.g., reading news can be more often done in the morning. From a very different perspective, time patterns might be very important in predicting the performance of restaurants or other physical venues~\cite{DBLP:journals/epjds/DSilvaNMMS18}. 

Another, use case for studying click streams is increasing the general performance of ad networks. On one hand, there are built models in order to detect and measure click-spam~\cite{click-spam}, or filter low-quality clicks that do not lead to conversions~\cite{10.1145/1566374.1566406}. On the other hand, one might want to address the problem of fraudulent pages, i.e., content farms~\cite{Luh2019IsIW}. However,~\cite{Vanessa2018ContextualMB} revealed that advertising on content farms might be still profitable. 

\section{Algorithm}

The main goal of the algorithm is to select web pages (domains) that are non-profitable and ads should be blocked from appearing on those web pages. The blocking might be restricted only to specific hours of the day. Due to the fact that conversions are very rare events and many domains have few clicks, we cannot wait for collecting sufficiently many clicks on every domain to estimate CR, hence we would like to analyse jointly some groups of domains and based on this build blocking rules (i.e., which clusters should be blocked completely or blocked during certain hours).

The algorithm consists of three steps:
\begin{enumerate}
    \item Clustering of domains based on their traffic-fingerprints.
    \item Computing statistics of clusters to assess their profitability and creating blocking rules.
    \item After each day computing updated vector representations of domains and reassigning domains to clusters, and finally blocking the domains from the selected clusters at the selected hours of the day.
\end{enumerate}

In such a way we are able to boost campaigns' performance by cutting off the low-quality non-profitable traffic that generates a lot of clicks but few conversions.

\subsection{Step 1 -- Clustering of domains}\label{clustering}

For each domain we count clicks $c_i$ for each hour of the day $i=0,\ldots,23$ over the whole history of our interaction with these domains in all our marketing campaigns. Then we normalize the values by the total number of clicks on the domain -- the resulting vectors are called traffic-fingerprints $f_i$ for $i=0,\ldots,23$ is defined as:
\[
f_i = \frac{c_i}{\sum_{j=0}^{23}c_j}.
\]
The next step is described

Next, we cluster such 24-dimensional vectors using the $k$-means algorithm.

For determining the optimal number of clusters, for each ad network we start with 2 clusters and then gradually increase the number as far as adding another cluster does not give a relevant improvement of a sum of differences between elements and clusters’ centers, according to the so-called Elbow method\footnote{https://en.wikipedia.org/wiki/Elbow\_method\_(clustering)}. 


Finally, we measure the clusters' quality using silhouette score~\cite{rousseeuw1987silhouettes}.

\subsection{Step 2 -- Creating blocking rules}




For each cluster we study the dependence of profit on the hour. 
The shape of these graphs looks similar for different networks, but for some, both clusters are above profitability around the clock, and for others, the weaker clusters "dive" below profitability for some part of the day. Using this dependence we generate blocking rules that contain hours with negative profit. It was always the case that the night hours are weaker, therefore it is possible to set two points - in the evening and in the morning - which are the limit of profitability.
Turning the campaign off and on for big clusters of domains is a time-consuming process and ad networks need some period of time to re-optimize campaigns due to changes in the campaign's configuration. Therefore, it only makes sense that it is done sparingly (e.g.~once per day). 

\subsection{Step 3 -- Reassigning domains to clusters}

Having computed clusters of domains on historical data, we fix clusters' centers (in the k-means algorithm each cluster is defined by its centroid). Then, while we are collecting new traffic on domains, we update traffic-fingerprints for domains (in the same way as in Step 1). Finally, for each domain we select the closest cluster (in the Euclidean metric). We decided to not redo clustering after new data is collected. Our offline simulations suggest that this does typically change the final blocking decisions significantly. However, it does require monitoring the number and the character of clusters found and matching them to the original ones which is an unnecessary complication.

What is important, when a domain is blocked during some hours on a given day, we do not include statistics for this day in the updated traffic-fingerprints. Otherwise, we would disturb the representation of this domain's natural traffic. 

\section{Offline experiments}


\paragraph{Experimental environment}

We performed experiments with our algorithm on several ad networks. For each ad network we split the data into train and test sets. The information about data sets and the split is given in Table~\ref{tab:data_stats}. For clustering, we selected only domains with at least 50 clicks to ensure the robustness of domains' vectors. On the train set we selected the optimum number of clusters and made decisions which clusters at which hours should be blocked (steps (1) and (2) of the algorithm). Then, we proceeded according to the chosen strategy on the test set - by recomputing traffic-fingerprints, reassigning domains to clusters (step (3)) and simulating clusters' blocking. Finally, we compared the statistics obtained in the simulation with the real statistics.

We should notice that in this approach the results can be inaccurate, because in reality blocking domains has an impact on the behaviour of ad networks, i.e., after blocking some domains we will observe additional traffic on other domains (because the ad network could aim to use the whole available budget). Hence, only online experiments (described in the next section) can give us definitive results. However, in the offline experiments we were able to test a wider range of traffic and check if the clusters' patterns identified in one ad network are similar to clusters in other ad networks (and that we avoided overfitting to a specific ad network's strategy).

\begin{table}[t]
\small
\centering

\caption{Basic statistics for obtained clusters in examined ad networks. Increase of profit shown here was obtained in offline test.}
\label{tab:offline_results}
\begin{tabular}{p{0.11\columnwidth}rrrr}
\toprule
\parbox{0.13\columnwidth}{Ad \mbox{network}} & \parbox{0.14\columnwidth}{Silhouette score} & \parbox{0.13\columnwidth}{Increase of profit} &  Cluster & Blocking rule \\
\midrule
MGID & 0.14 & 41\% & 0 & Blocked 16 - 24 \\
 && & 1 & Not blocked \\
 && & 2 & Blocked \\
 && & 3 & Blocked \\
\hline
Exoclick & 0.18 & 211.3\% & 0 & Not blocked \\
 && & 1 & Blocked 20 - 5 \\
\hline
\multirow{2}{*}{\parbox{0.13\columnwidth}{Content Stream}}& 0.19 & 4.9\% & 0 & Blocked \\
 && & 1 & Non blocked \\
\hline
Taboola & 0.14 & 7.6\% & 0 & Not blocked \\
 && & 1 & Blocked  15 - 5 \\
\hline
\multirow{2}{*}{\parbox{0.13\columnwidth}{Traffic Stars}} & 0.12 & 581.1\% & 0 & Blocked 22 - 6 \\
 && & 1 & Blocked \\

\bottomrule
\end{tabular}

\end{table}

\begin{figure}
     \centering
     \begin{subfigure}[b]{0.4\textwidth}
         \centering
         \includegraphics[width=\textwidth]{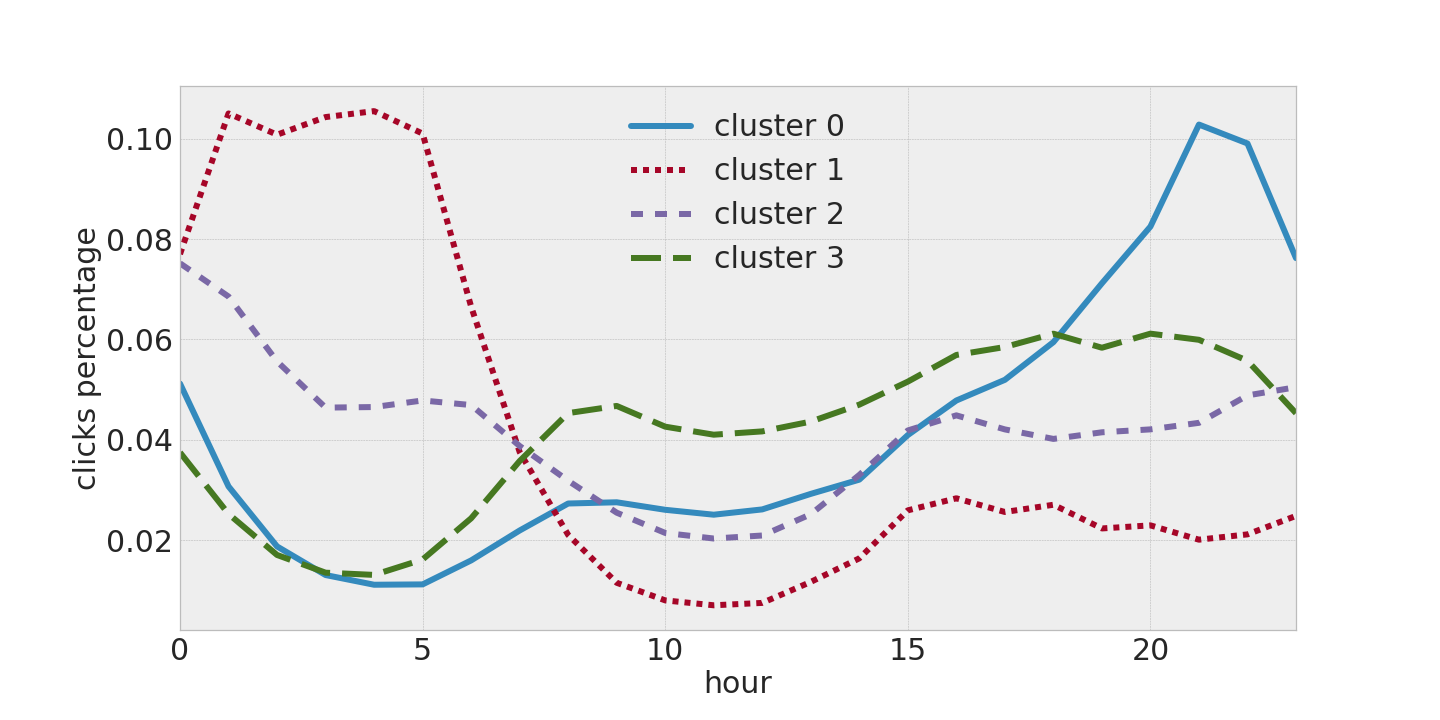}
         \caption{MGID}
         \label{fig:mgid_clusters}
     \end{subfigure}
     \hfill
     \begin{subfigure}[b]{0.4\textwidth}
         \centering
         \includegraphics[width=\textwidth]{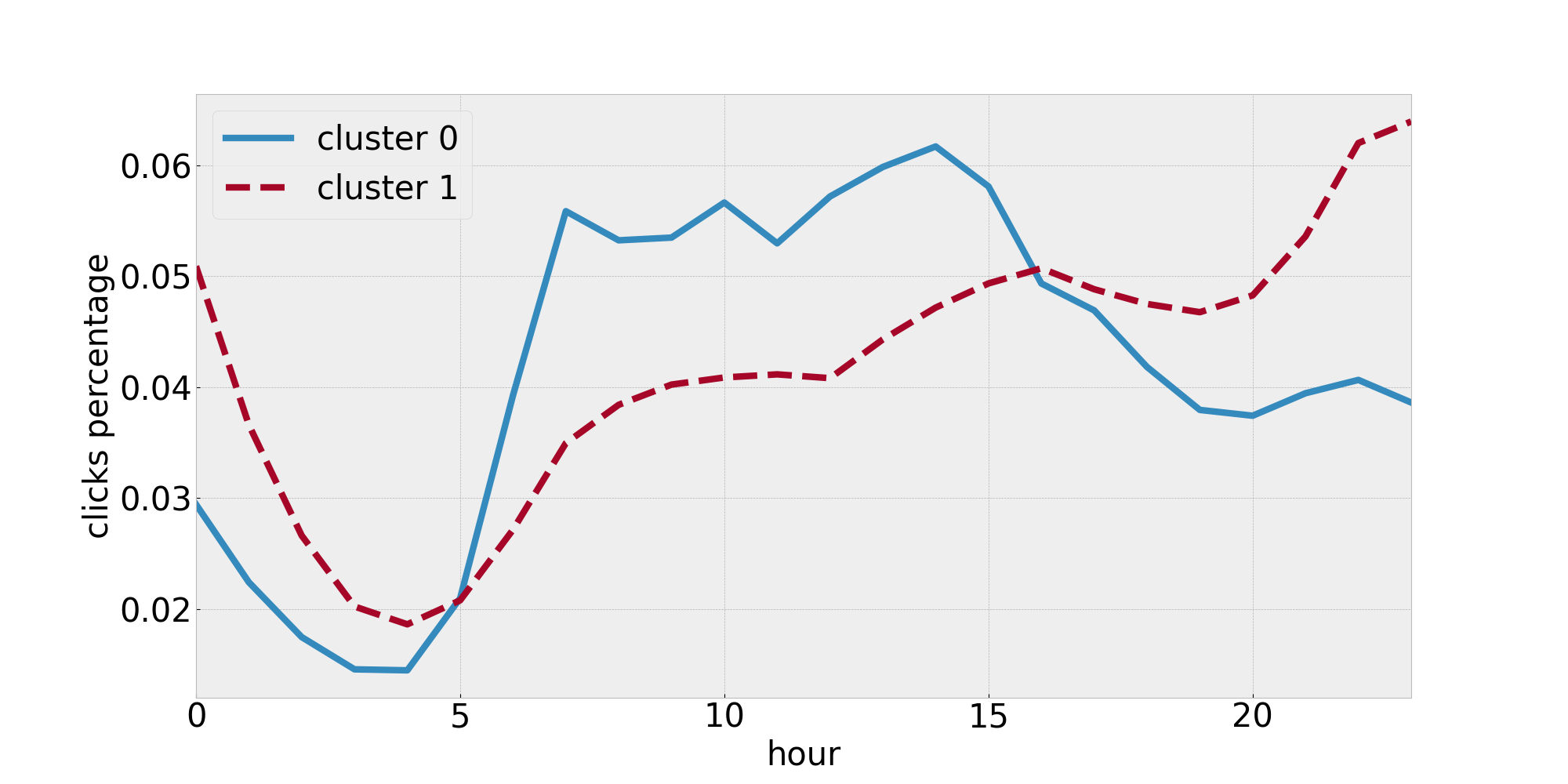}
         \caption{Exoclick}
         \label{fig:exoclick_clusters}
     \end{subfigure}
     \hfill
     \begin{subfigure}[b]{0.4\textwidth}
         \centering
         \includegraphics[width=\textwidth]{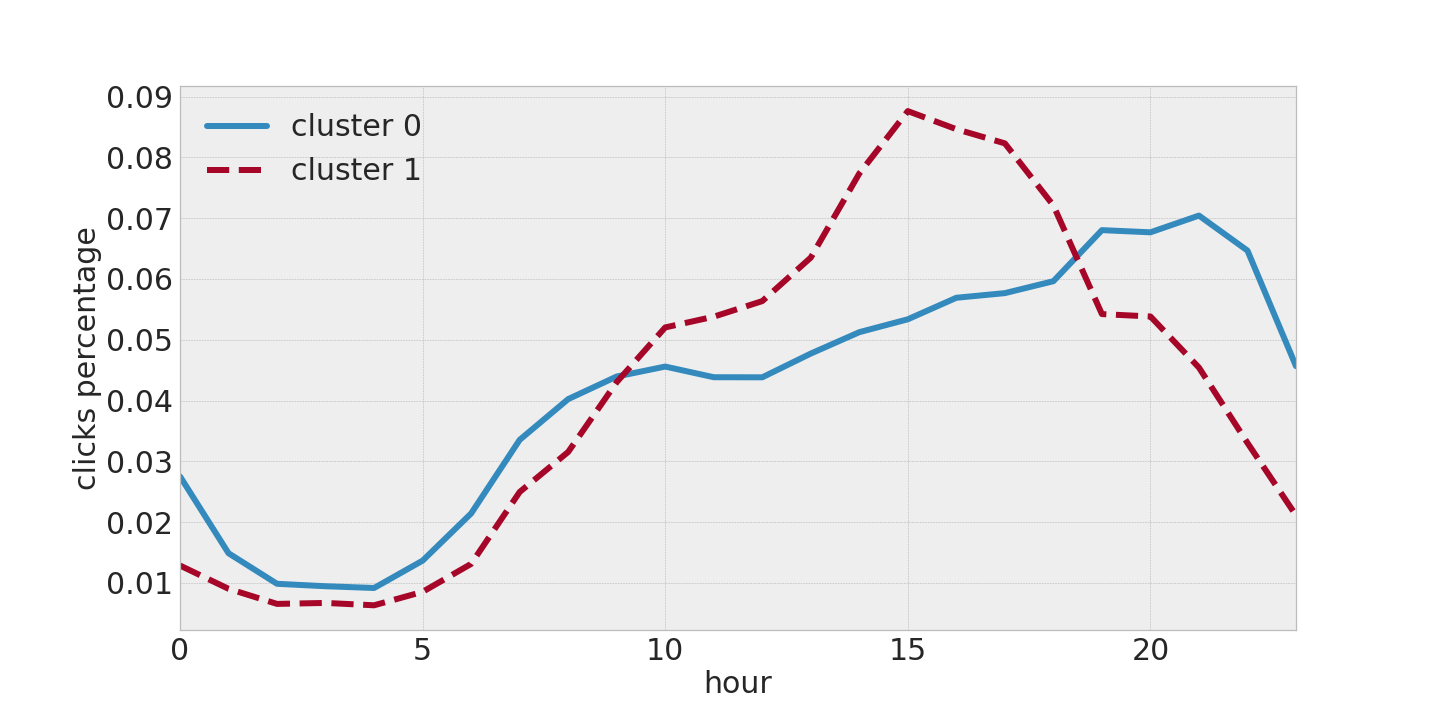}
         \caption{Content Stream}
         \label{fig:content_stream_clusters}
     \end{subfigure}
     \hfill
     \begin{subfigure}[b]{0.4\textwidth}
         \centering
         \includegraphics[width=\textwidth]{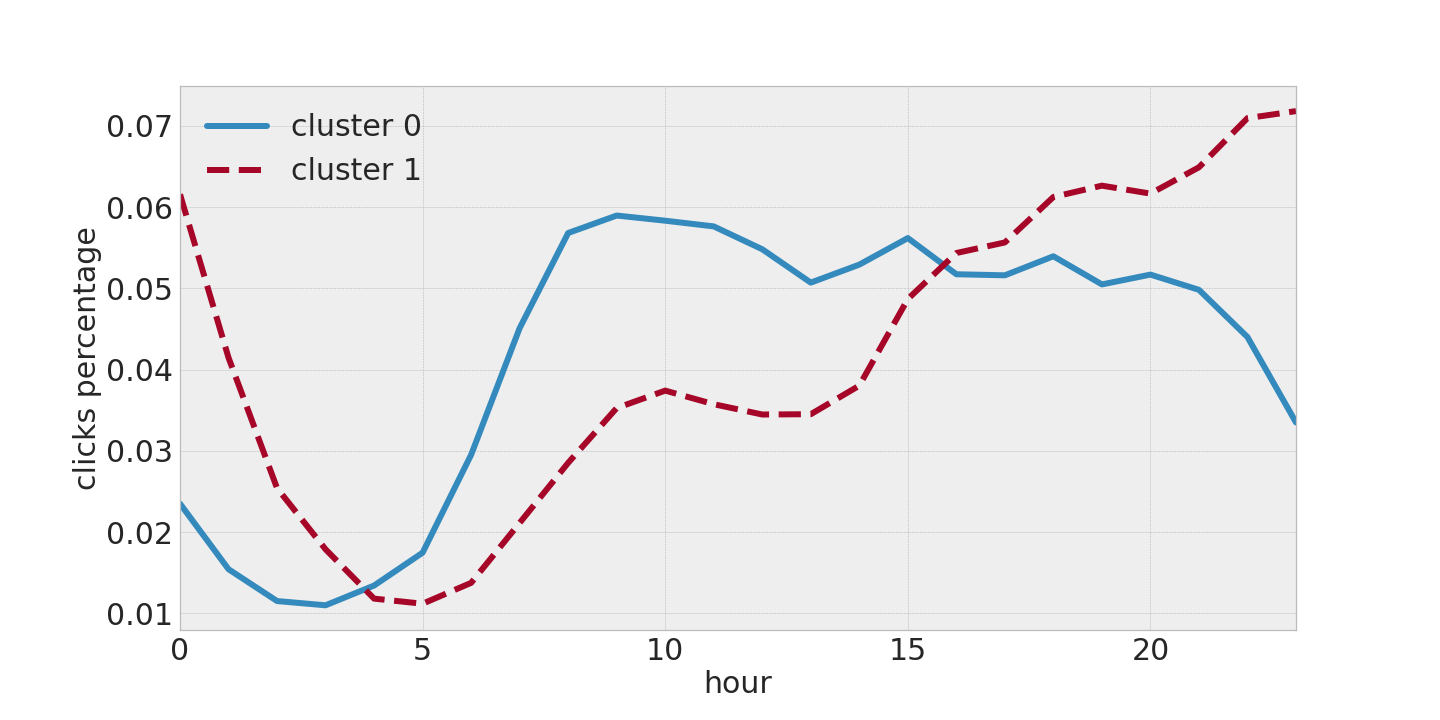}
         \caption{Taboola}
         \label{fig:taboola_clusters}
     \end{subfigure}
     \hfill
     \begin{subfigure}[b]{0.4\textwidth}
         \centering
         \includegraphics[width=\textwidth]{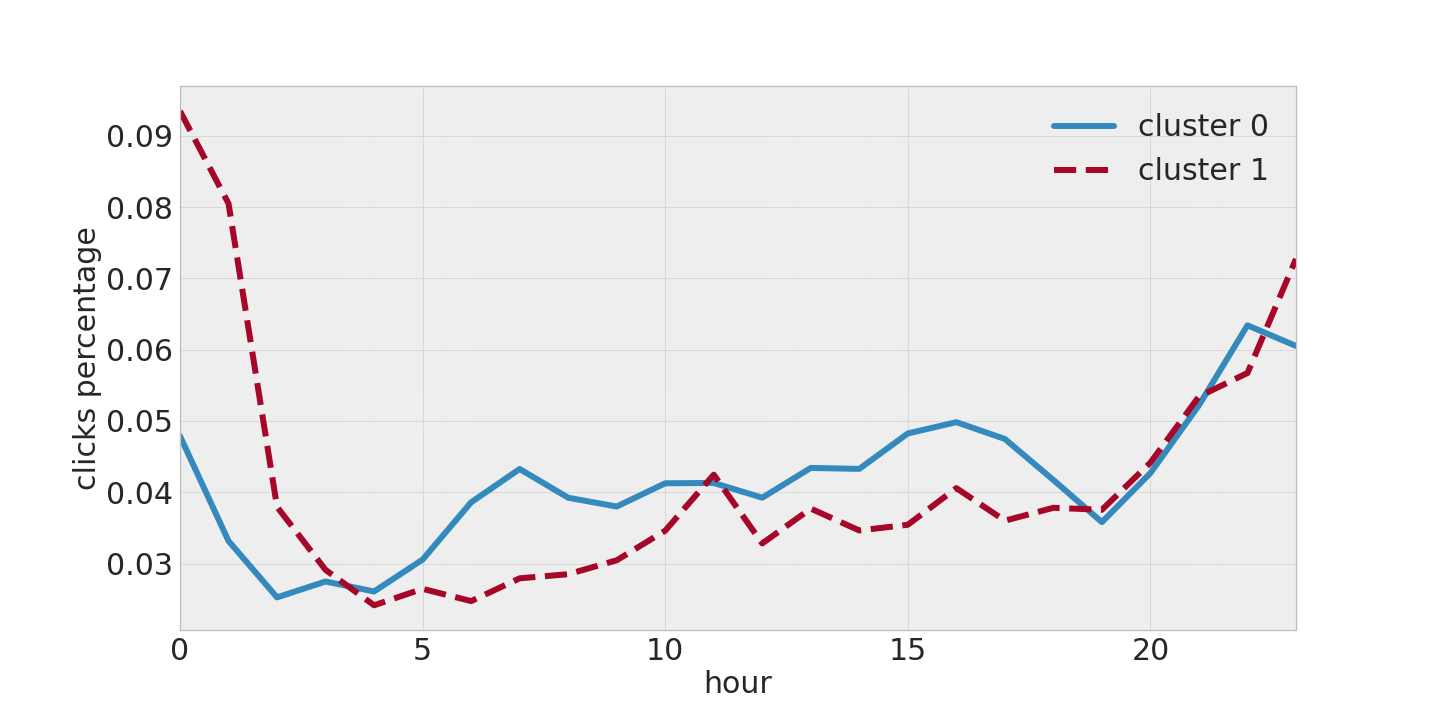}
         \caption{Traffic Stars}
         \label{fig:traffic_stars_clusters}
     \end{subfigure}
        \caption{Traffic-fingerprints for identified clusters' centers for studied ad networks. A clear difference of daily traffic patterns is seen between the clusters. }
        \label{fig:clusters_centers}
\end{figure}

\begin{figure}[]
    \centering
    \includegraphics[width=9cm]{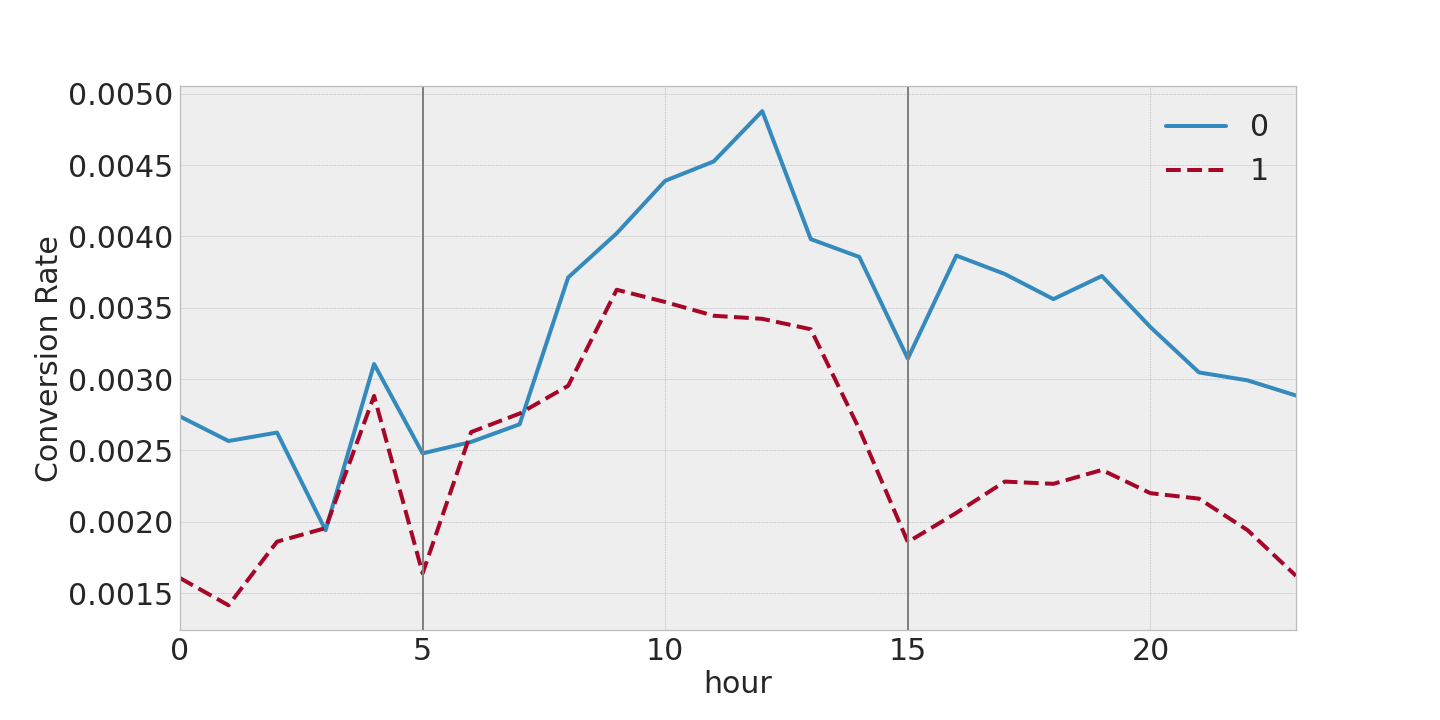}
    \caption{Conversion rate, in dependence on hour of day, for two clusters of Taboola clustering. Vertical lines show hours of (un)blocking cluster 1.}
    \label{fig:taboola_CR}
\end{figure}

\begin{table}[t]
\footnotesize
\centering

\caption{Split of data into train and test sets.}
\label{tab:data_stats}
\begin{tabular}{lr}
\toprule
Ad network &  Start, split, end date \\
\midrule
MGID & 2020-03-16, 2021-04-23, 2021-06-30 \\
\hline
Exoclick & 2019-01-01, 2021-05-21, 2021-06-20 \\
\hline
Content Stream& 2020-01-01, 2021-12-28, 2022-01-26 \\
\hline
Taboola & 2020-01-01, 2021-12-28, 2022-01-26  \\
\hline
Traffic Stars & 2020-01-01, 2021-12-28, 2022-01-26\\

\bottomrule
\end{tabular}

\end{table}

\begin{table}[t]
\footnotesize
\centering

\caption{Information about train and test sets used in offline experiments.}
\label{tab:data_stats}
\begin{tabular}{lrrrr}
\toprule
Ad network &
Budget (train / test) & Clicks (train / test) & Domains (train / test) \\
\midrule
MGID &  54,767 / 18,997 & 5,416,140 / 825,672 & 304,996 / 62,036 \\
\hline
Exoclick &  1,332,013 / 35,436 & 41,483,870 / 1,524,199
& 22,052 / 5,938\\
\hline
Content Stream& 218,870 / 16,635 & 4,029,076 / 427,148 & 2,246 / 732 \\
\hline
Taboola &  295,327 / 13,810 & 4,542,487 / 223,299 &
8,540 / 1,434 \\
\hline
Traffic Stars & 195,445 / 1,621 & 6,331,319 / 58,005 & 1,630 / 807 \\

\bottomrule
\end{tabular}

\end{table}

\paragraph{Results}

Table~\ref{tab:offline_results} shows the basic statistics obtained in offline experiments for five examined ad networks. We can see the selected number of clusters with the silhouette score and the rules for blocking that were obtained on the training set. The column \textit{Increase of profit} shows us the gain after applying the rules on the test set. For every ad network the increase was positive, however, with a high variance between the obtained results.

For most cases the optimum number of clusters was 2, from which one was profitable and the second -- non-profitable for at least a part of a day. On Figure~\ref{fig:clusters_centers} we can see daily patterns of clusters' centers. It turned out that in general the cluster that has more traffic during evening and night hours is less profitable.

A very interesting observation is that the less profitable clusters have lower CR not only during evening and night hours, but also for other hours of the day. The Figure~\ref{fig:taboola_CR} illustrates this phenomenon for Taboola clustering, but for other ad networks we could observe similar patterns. This indicates that the identified clusters are not 
just some coincidence, but represent domains with a different type of daily activities
that can be identified by their traffic and thus have a different value for e-commerce.

We have suspected that the revealed clusters might correspond to some topical groups of pages. Thus we have investigated contextual data delivered the networks which make this data available, i.e., all but MGID. We were unable to see any topical characteristics of the clusters. In our case, these networks did provide us with very homogeneous traffic (e.g., Content Stream delivered only news-like web pages), and we were not able to spot any significant topical difference between the two clusters used by our algorithm.

\section{Online experiments}


We executed online experiments with our algorithm on several campaigns on the MGID platform. The campaign parameters were configured by human operators who selected such values as a market, advertisements, appropriate stakes per ad and so on. Then during the whole examined period the operators have been monitoring the campaigns' profitability and have been performing necessary actions such as modifying stakes or changing ads. The operators based their decisions on their experience and simple rules like "if the ad is not profitable, remove it and add a new one with a higher stake".

When we introduced the algorithm for blocking clusters of domains, the operators have been acting on the campaigns as before. Hence, the campaigns were optimized semi-automatically, but still the final result of the campaigns depended highly on the correctness of the human actions.

The exact A/B tests were impossible, because the traffic for our campaigns is provided by the ad network and we are not able to directly choose the web pages, where we will display ads. So we are not able to create two exactly the same campaigns.

Hence, we decided to select a subset of all active campaigns and to run the algorithm on this subset (a similar approach for an experimental design was proposed e.g. in~\cite{agarwal2014budget}), while the other part of campaigns was managed as before by human operators. This way, we could compare the periods before and after running optimization, and also the results for two campaign groups.

We have been monitoring all campaigns' performance on MGID from the beginning of June 2021 (the number of active campaigns varied before 20 and 40, since the campaigns were constantly paused and restarted by the operators due to market changes). Then, on the 7th of July we randomly selected about half of the campaigns and began the optimization by the algorithm. Finally, on the 9th of September we turned on the algorithm on all active campaigns.

\begin{figure}[]
    \centering
    \includegraphics[width=9cm]{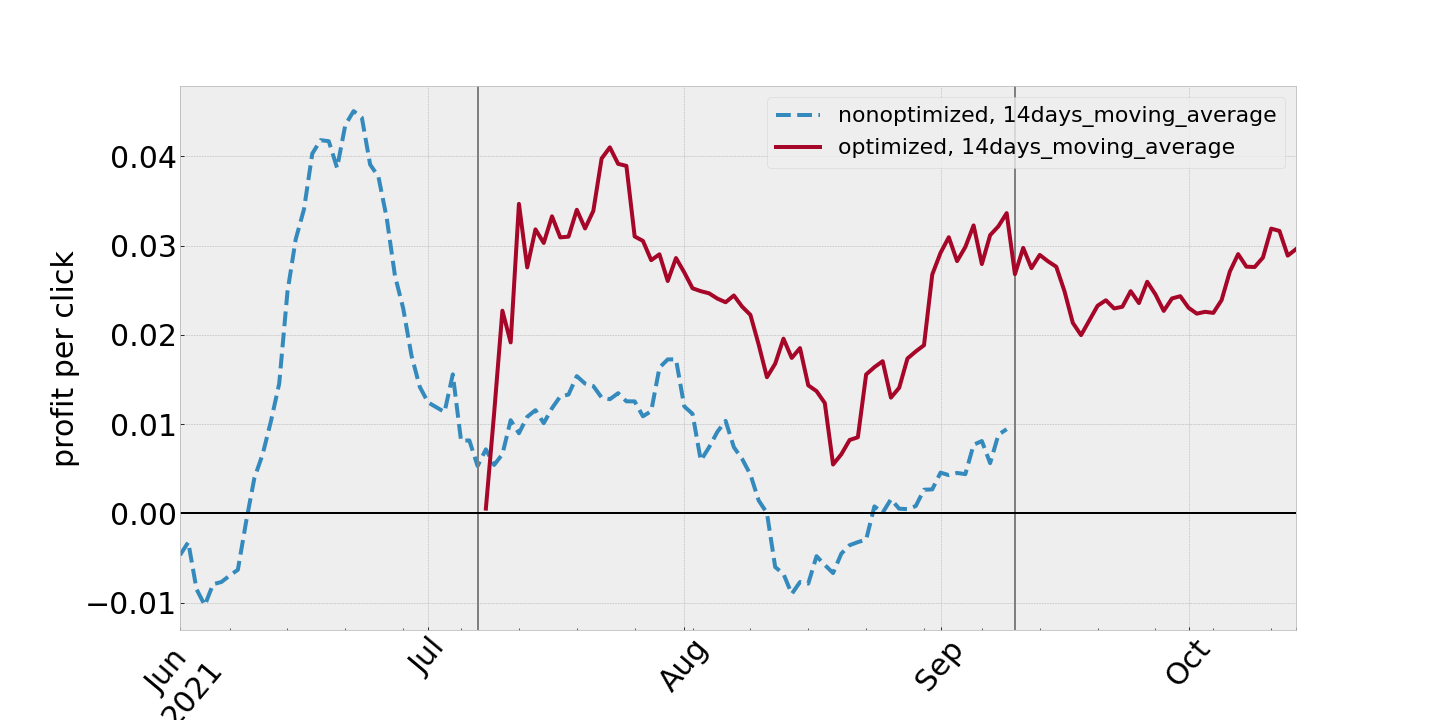}
    \caption{Average profit per click for all campaigns in A/B tests. The dashed line shows the performance of the system without automatic cluster blocking, whereas the solid line shows results with it. Vertical lines show the days when the algorithm was turned on for some campaigns: 7th of July half of active campaigns and 9th of September all campaigns.}
    \label{fig:ppc}
\end{figure}

Figure~\ref{fig:ppc} shows 14-days moving averages of the total profit per click for optimized and non-optimized campaigns. Vertical lines indicate the days when the algorithm was turned on for some campaign groups. Despite high variability of the metric, we can observe a noticeable jump in the performance after turning on the optimization in both time stamps.

In Table~\ref{tab:14days} we can see the other metrics for 37 campaigns that were optimized for 14 days only by operators and 14 days later also by the algorithm. After turning on the algorithm, total profit increased by about 53.4\%, while the profit per click increased by 37\%. What is important, despite temporary blocking some web pages, the total volume of the traffic has not decreased (actually, the number of clicks even slightly increased). 

\begin{table}[t]
\centering
\small
\caption{Statistics for 14 day periods before and after blocking was introduced. The profit increased by more than 50\%.}
\label{tab:14days}
\begin{tabular}{lrrrrrr}
\toprule
\parbox{0.13\columnwidth}{Blocked period}&  Clicks &  \parbox{0.11\columnwidth}{Conver\-sions} &  CR \% &  Profit &  ROAS &  \parbox{0.13\columnwidth}{Profit per click} \\
\midrule
False              &  186,528 &          760 &  0.41 &     3931.55 &  1.23 &                  0.021 \\
True               &  208,773 &          905 &  0.43 &     6030.20 &  1.31 &                  0.029 \\
\bottomrule
\end{tabular}

\end{table}

\section{Conclusions}

We demonstrated that in order to effectively manage blocking domains in ad networks, it is enough to collect just 50 clicks per single domain. The proposed domains' blocking algorithm allowed us to increase profit by over 40\% both in offline simulations on historical data and as well as online on real e-commerce campaigns. We suspect that such improvement was not caused by traffic management algorithms in the MGID ad network, but is a result of real differences in traffic quality between domains. This is testified by similar shapes of clusters' centroids traffic-fingerprints in all considered ad networks.

In the future work we plan to extend our approach as follows: training supervised learning models, that using domains' traffic-fingerprints predict domain profitability. Next, use these models in order to block domains predicted as unprofitable. However, this approach will have many challenges, like interpretability and building blocking rules that do not turn on and turn off domains too often. In this aspects the proposed clustering method is more understandable from a business point of view and easier to use. 

Furthermore, our current solution automates only partially the tasks necessary to run profitable e-commerce online campaigns. The next steps should focus on the full substitute of human campaign operators by sets of algorithms. Still, the presented solutions would form one of the most innovative aspects of the system we develop.

\section{Acknowledments}

This work was supported by The National Centre for Research and Development (NCBiR) grant no. POIR.01.01.01-00-0945/19.

\bibliographystyle{splncs04}
\bibliography{sample-base}

\end{document}